\documentclass[article]{aa}  
\usepackage{graphicx}
\usepackage{xcolor}
\usepackage{lscape}
\usepackage{natbib}
\usepackage{multirow}
\usepackage{txfonts}
\bibpunct{(}{)}{;}{a}{}{,} 

\newcommand{\emm}[1]{\ensuremath{#1}}
\newcommand{\emr}[1]{\emm{\mathrm{#1}}}

\newcommand{\unit}[1]{\emr{\,#1}}

\newcommand{\kms}{\unit{km~s^{-1}}}

\newcommand{\arcs}{$^{\prime\prime}$}

\begin{document}

\title{Evidence for disks at an early stage in class 0 protostars? }

\author{M. Gerin \inst{1} %
  \and J. Pety \inst{2,1} %
  \and B. Commer\c con \inst{3} %
  \and A. Fuente \inst{4} %
  \and J. Cernicharo \inst{5} %
  \and N. Marcelino \inst{5} %
  \and A. Ciardi \inst{1} %
  \and D. C. Lis \inst{6} %
  \and E. Roueff \inst{1} %
  \and H.A. Wootten \inst{7} %
  \and E. Chapillon \inst{8,2} %
}

\institute{%
  LERMA, Observatoire de Paris, PSL Research University, CNRS, Ecole
  Normale Sup\'erieure, Sorbonne Universit\'es, UPMC Univ. Paris 06,
  F-75005 Paris, France.  \email{maryvonne.gerin@ens.fr} %
  \and Institut de Radioastronomie Millim\'etrique (IRAM), 300 rue de la
  Piscine, 38406 Saint Martin d'H\`eres, France. %
  \and Univ Lyon, Ens de Lyon, Univ Lyon1, CNRS, Centre de Recherche
  Astrophysique de Lyon UMR5574, F-69007, Lyon, France. %
  \and Observatorio Astron\'omico Nacional (OAN,IGN), Apdo 112, E-28803
  Alcal\'a de Henares, Spain. %
  \and Instituto de Ciencia de Materiales de Madrid (ICMM-CSIC). E-28049,
  Cantoblanco, Madrid, Spain. %
  \and LERMA, Observatoire de Paris, PSL Research University, CNRS,
  Sorbonne Universit\'es, UPMC Univ. Paris 06, F-75014 Paris, France. %
  \and National Radio Astronomy Observatory, 520 Edgemont Road,
  Charlottesville, VA 22903, USA. %
  \and LERMA, Observatoire de Paris, PSL Research University, CNRS,
  Sorbonne Universit\'es, UPMC Univ. Paris 06, Ecole Normale Sup\'erieure,
  F-92190 Meudon, France. %
  \and Laboratoire d'astrophysique de Bordeaux, Univ. Bordeaux, CNRS, B18N,
  all\'ee Geoffroy Saint-Hilaire, 33615 Pessac, France. %
}

\date{Received xxx; accepted xxx}
 
\abstract {}
{The formation epoch of protostellar disks is debated because of the
  competing roles of rotation, turbulence, and magnetic fields in the early
  stages of low-mass star formation.
  Magnetohydrodynamics simulations of collapsing cores predict that
  rotationally supported disks may form in strongly magnetized cores
  through ambipolar diffusion or misalignment between the rotation axis and
  the magnetic field orientation.  Detailed studies of individual sources
  are needed to cross check the theoretical predictions.}
{We present 0.06 -- 0.1 \arcs\ resolution images at 350~GHz toward B1b-N
  and B1b-S, which are young class 0 protostars, possibly first hydrostatic
  cores. The images have been obtained with ALMA, and we compare these data
  with magnetohydrodynamics simulations of a collapsing turbulent and
  magnetized core. }
{The submillimeter continuum emission is spatially resolved by ALMA.
  Compact structures with optically thick 350~GHz emission are detected
  toward both B1b-N and B1b-S, with 0.2 and 0.35\arcs\ radii (46 and 80~au
  at the Perseus distance of 230~pc), within a more extended envelope.  The
  flux ratio between the compact structure and the envelope is lower in
  B1b-N than in B1b-S, in agreement with its earlier evolutionary
  status. The size and orientation of the compact structure are consistent
  with 0.2\arcs\ resolution 32~GHz observations obtained with the Very
  Large Array as a part of the VANDAM survey, suggesting that grains have
  grown through coagulation.  The morphology, temperature, and densities of
  the compact structures are consistent with those of disks formed in
  numerical simulations of collapsing cores. Moreover, the properties of
  B1b-N are consistent with those of a very young protostar, possibly a
  first hydrostatic core.  These observations provide support for the early
  formation of disks around low-mass protostars.}
{}


\maketitle
%

\section{Introduction}
A key issue in the early stages of star formation is the formation epoch of
a rotationally supported disk, which mediates the accretion onto the
forming star and evolves into a protoplanetary disk at later stages.
Analytical and numerical studies have shown that the formation of a
rotationally supported disk requires leaving enough angular momentum during
the collapse. This can be a problem when a magnetic field is present
because the field can efficiently transport angular momentum and prohibits
the formation of a disk \citep[e.g.,][]{hennebelle:08}.  Earlier
theoretical studies were mostly restricted to the simple case of a magnetic
field parallel to the rotation axis, a configuration where magnetic braking
can be very efficient and quench the formation of a disk. The B335
protostar seems to be the best case of such a geometry \citep{yen:15}. The
occurrence frequency of such systems is unknown, however. They may be the
exception rather than the rule, since \citet{lee:16} have shown that
outflow axes of wide binary or multiple systems are not aligned in Perseus,
suggesting that these systems are formed in more complex configurations.
Different effects contribute to reducing the efficiency of magnetic braking
and allowing early formation of disks: a rotation axis not aligned with the
magnetic field \citep{ciardi:10}, the level of turbulence \citep{joos:13},
or deviations from ideal magnetohydrodynamics (MHD), especially the role of
ambipolar diffusion \citep{tomida:15,masson:15,zhao:16,hennebelle:16}.

Disks have been commonly found around class I protostars and class II
sources \citep{williams:11,harsono:14}, but their presence around younger
objects, especially class 0 protostars, is still debated. Establishing the
presence of rotationally supported disks in class 0 protostars
\citep[defined as protostars with a high ratio of the submillimeter to the
bolometric luminosity;][]{andre:93} is especially important because objects
at this stage are accreting material from their surrounding envelope at the
highest rate. Therefore these objects are best suited for a comparison with
theoretical collapse models, especially regarding the question of the
complex interplay of velocity field and magnetic field on the final
structure of the environment of the forming star.  The number of such
protostars that can be studied in detail is limited, and direct high
angular resolution imaging in the near-infrared is often impossible because
of the very high column densities of dust and gas in the surrounding
envelopes. Only (sub)millimeter interferometers have the sensitivity and
angular resolution to detect circumstellar material.  \citet{maury:10}
performed a pilot study of multiplicity with the IRAM-PdBI at $\sim 0.4''$
resolution that showed a low multiplicity fraction at small separations.
At the achieved angular resolution, the class 0 protostars are detected,
but the structure of the millimeter emission is not resolved.  The young
protostars B1b-S and B1b-N are promising objects for studying the early
stages of low-mass star formation. These sources are located in the
\object{Barnard~1b} core of the \object{Barnard~1} dark cloud, a moderately
active star-forming region in the Perseus molecular cloud at a distance of
230 pc \citep{hirota}.  The analysis of the spectral energy distribution
(SED) based on {\sl Herschel} and {\sl Spitzer} data performed by
\citet{pezzuto:12} shows that both sources have low infrared luminosities
($L_\mathrm{bol}$ = 0.1 L$_\odot$ for B1b-N $L_\mathrm{bol}$ = 0.3
L$_\odot$ for B1b-S), with the SED peaking longward of 100 $\mu$m.  To
further progress in the understanding of the nature of these sources and
their relation to their parental core, \citet{hirano:14} obtained CO ($J=2
\rightarrow 1$), $^{13}$CO ($J=2 \rightarrow 1$) and H$^{13}$CO$^+$ ($J=1
\rightarrow 0$) observations with the SMA, which showed that both
\object{B1b-S} and \object{B1b-N} are driving low-velocity molecular
outflows. The continuum sources are largely unresolved down to an angular
resolution of 0.5\arcs\ at 7~mm wavelength \citep{hirano:14}.
\citet{huang:13} showed that N$_2$H$^+$ ($J=3\rightarrow 2$) and N$_2$D$^+$
($J=3\rightarrow 2$) present a compact emission component associated with
the continuum sources, together with extended emission in the surrounding
envelope. The high intensity of these high-dipole moment species has been
interpreted by \citet{daniel:13} using a sophisticated radiative transfer
code. As expected for a cold and dense region, the deuterium fractionation
clearly increases with the density, although the absolute abundances slowly
decrease.  Using NOEMA with a 2.3\arcs\ beam, \citet{gerin:15} further
confirmed the presence of slow molecular outflows driven by B1b-N and
B1b-S, and determined the outflow dynamical times, masses, mechanical
luminosities, and mass-loss rates.  They concluded that the two sources are
at most a few thousand years old, and that the properties of B1b-N are
consistent with those of a first hydrostatic core (FHSC). The outflows are
launched in different directions, which is probably a consequence of the
high degree of turbulence in the Barnard~1b core, which is impacted by
outflows from nearby young stellar objects (YSOs) \citep{hirano:14}.  So
far, the inner structure of the compact components of B1b-N and B1b-S
remained unresolved or only marginally resolved. Because these sources are
young and the physical and chemical structure of the Barnard~1b core are
well known \citep[see the recent analysis of the depletion and ionization
rate by ][]{fuente:16}, it is interesting to probe these sources in more
detail, as a means to constrain current theoretical understanding of
low-mass star formation.

\section{ALMA observations}
\begin{table*}
  \caption{Summary of combined observations}
  \label{tab:sources}
  \begin{tabular}{lccccccccc}
    \hline
    \hline
    Source & RA & Dec & Beam & $PA$$^a$ & $F_{max}^{DSB}$$^b$ &   $\sigma F$$^c$ & $F(1'')^{DSB}$$^d$ & $S_{32}^{350}(0.3'')$$^e$ \\ 
    & (J2000)   & (J2000)    & \arcs   & $^\circ$ & mJy/beam  & mJy/beam & Jy & \\
    \hline
    B1b-N & 03:33:21.209 & 31:07:43.66 &$0.086\times0.069$& 1.1& 13.4 & 0.71 &$0.73\pm 0.07$& $2.4\pm 0.1$\\
    B1b-S & 03:33:21.355 & 31:07:26.37 &  $0.14\times0.11$& 16 & 63.1  & 2.4 & $0.89\pm0.08$ & $2.8\pm0.1$\\
    \hline
    J0336+3218 & 03:36:30.108 & 32:18:29.34 & $0.065\times0.045$ & 180 & 356  & 0.4 & 0.35&\\
    J0359+3220 & 03:59:44.913 & 32:20:47.16 &  $0.064\times0.048$ & 180 & 150  & 0.3 & 0.15&\\
    \hline
  \end{tabular}
  \tablefoot{$^a$ Beam position angle, $^b$ peak flux density using both sidebands.  $^c$  rms noise level, $^d$ flux density integrated in a 1\arcs\ radius circular aperture, $^e$  spectral index 
    between 32 and 350~GHz in a 0.3\arcs\ radius circular aperture.  }
\end{table*}

We present ALMA Cycle 3 observations, performed on 2015 November 22 during
the Long Baseline Campaign, combined with supplementary Director's
Discretionay Time (DDT) observations obtained on 2016 June 16.  For the
first data set, a total of 46 antennas were used in the C36-7
configuration, with baselines from 82~m up to 15~km (projected on source),
providing an angular resolution of $\sim0.06''$, corresponding to $\sim
13$~au at the distance of Perseus, 230~pc.  Four continuum base-bands with
2 GHz bandwidth, 128 channels, and dual polarization received the digitized
signals from the Band 7 receiver.  Central sky frequencies in each band are
336.5 and 338.4 GHz in the lower sideband (LSB), and 348.5 and 350.5 GHz in
the upper sideband (USB). The channel spacing is 15.6~MHz, resulting in a
spectral resolution of $\sim$27 km\,s$^{-1}$ (Hanning smoothed).  Imaging
with the first dataset revealed artifacts attributable to a lack of shorter
spacings. For the second data set, a total of 38 antennas were used in the
C40-4 configuration with baselines from 15~m to 704~m. The setups were
otherwise identical.

Observations for both datasets were obtained in single sessions, providing
30 minutes on source combined.  The two fields B1b-N and B1b-S were
observed alternately for 55 seconds each, with a phase calibrator scan in
between, observed for 20 seconds.  The phase calibrator was
\object{J0336+3218}, located at an angular separation of $\sim$1.36$^\circ$
from the B1b cores.  A second calibrator (\object{J0359+3220}, located at
5.7$^\circ$ distance), observed every 5 min, was also included in order to
check the quality of the phase transfer (see Appendix \ref{sec:cal} for
more information).  Standard calibration was performed in CASA 4.5.0, while
for imaging and data analysis we used the GILDAS software and CASA
4.7.0. For the first dataset, two other quasars were used for flux
(J0238+1636) and bandpass (J0237+2848) calibration. The adopted flux
density of J0238+1636 is 1.06 Jy at 343.5 GHz.  For the second dataset,
Ceres was observed for flux calibration and J0237+2848 for bandpass
calibration. J0238+1636 and J0237+2848 are both among the regularly
monitored ALMA calibrators, and the obtained fluxes are consistent with the
data in the ALMA Calibrator Database. For the DDT observations, the Ceres
data could not be used for flux calibration and J0237+2848 (flux 1.10 Jy at
343.5 GHz) was used instead.  Weather conditions were good and appropriate
for Band 7 observations, with a precipitable water vapor of 0.8 (0.7) mm
and a median system temperature of 170 (160) K for the long-baseline (DDT)
sessions.

The flux calibration we achieved is very good, with the curves of flux
versus $uv$ distance overlapping for the two program sources, as
illustrated in Fig.~\ref{fig:uv} (although the gain calibrator's flux
varied somewhat).  Extended flux is clearly present for both sources, as
the recovered flux increases with decreasing baseline length. Spectral
windows for each sideband were averaged for the combined data.

For the high-resolution data, the final clean beam is $0.064'' \times
0.048''$ at PA $168^\circ$with a noise level of 0.1~mJy/beam for each
spectral window, corresponding to about 0.35~K in the LSB and 0.32~K in the
USB.  For the DDT data, the final clean beam is $0.39'' \times 0.27''$ at
PA $-9.9^\circ$, with a noise level of 0.2~mJy/beam, corresponding to about
18~mK.

We have checked that all spectral windows are free of strong line emission.
With the typical line widths in Barnard~1b of $\sim 1$~km\,s$^{-1}$, which
can increase to $\sim 5$~km\,s$^{-1}$ when line wings are present, line
emission will be diluted in the broad spectral channels used for continuum
observations.  Although we cannot exclude some line contamination, we are
confident that the contribution of spectral lines to the detected continuum
fluxes remains modest.

For the data reported here, we have concatenated the two datasets. The
final clean beams are $0.086'' \times 0.069''$ for B1b-N, and $0.14''
\times 0.11''$ for B1b-S. We averaged the two spectral windows of each
sideband to obtain the final images.

\section{Results}
\label{sec:results}

\begin{figure*}
  \includegraphics[width=0.44\textwidth]{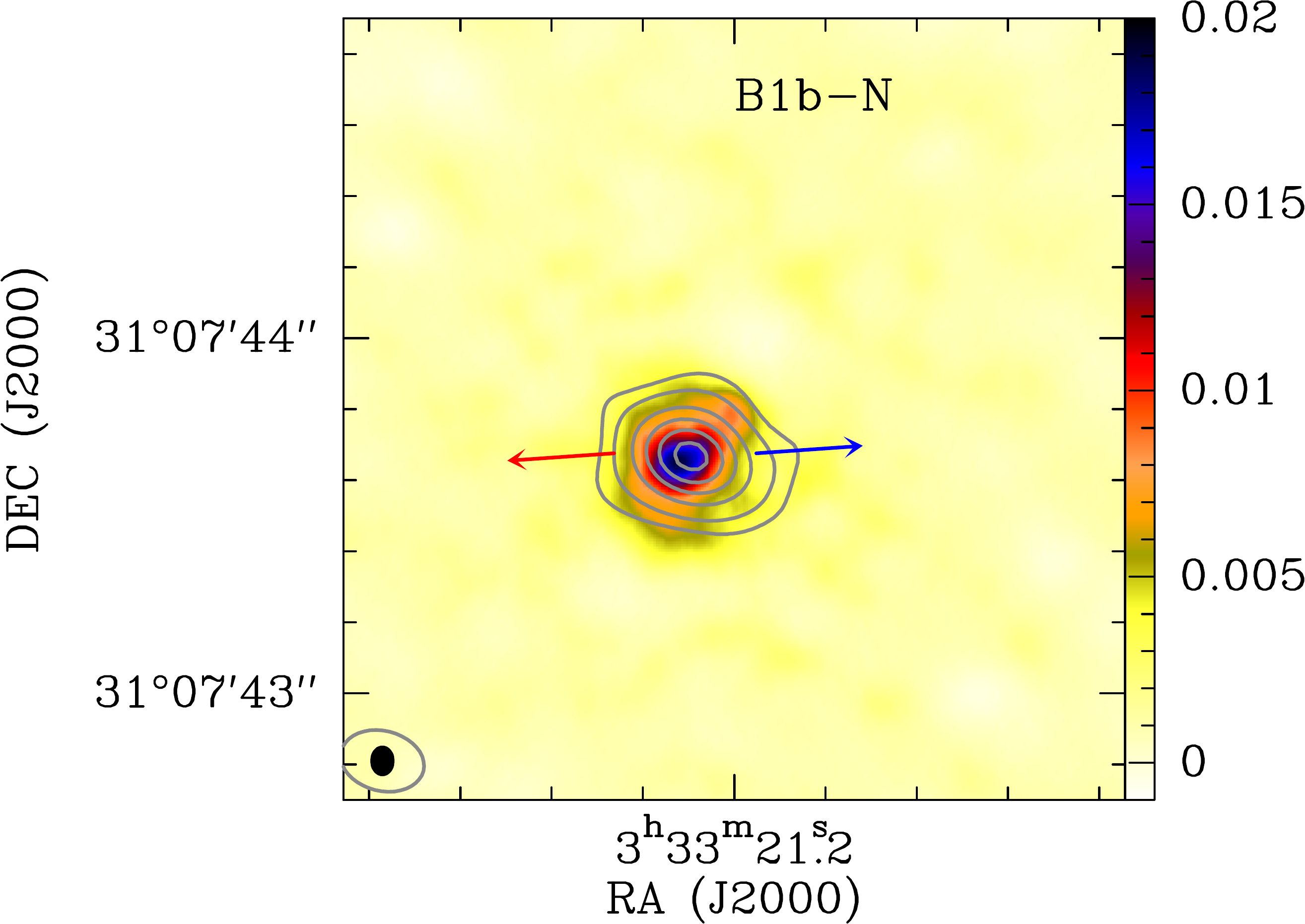}
  \hspace*{0.1cm} \includegraphics[width=0.4\textwidth]{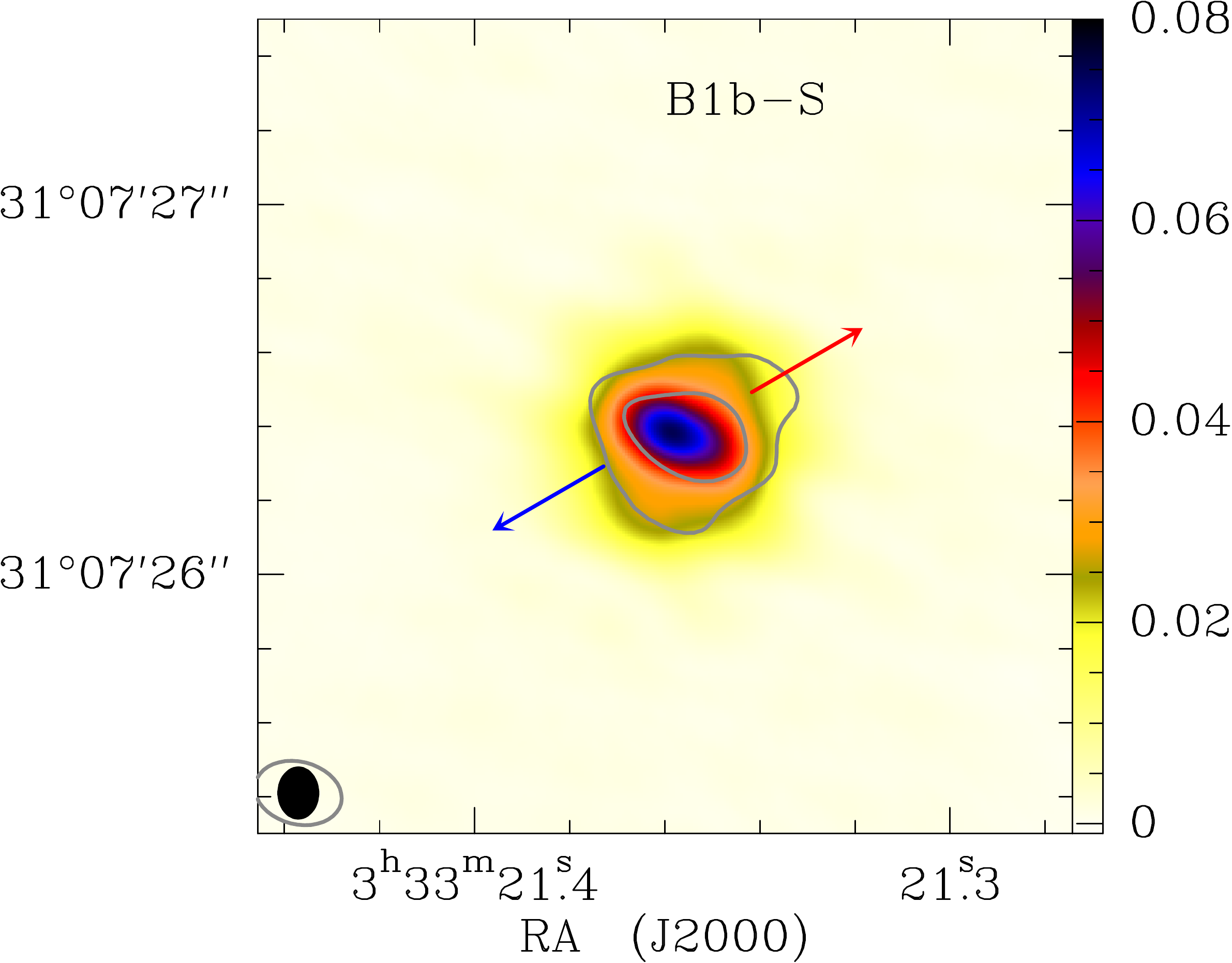}
  \caption{Images of the continuum emission at 349~GHz toward B1b-N (left)
    and B1b-S (right).  The gray contours show the 32~GHz continuum
    emission from the VANDAM survey \citep{tobin:15a,tobin:16} and are
    drawn at 0.05, 0.1, 0.2, 0.3, 0.4, and 0.5 mJy/beam. The ALMA and VLA
    beam sizes are shown as black and gray ellipses. The red and blue
    arrows show the approximate direction of the outflows.}
  \label{Fig:sources}
\end{figure*}

\subsection{Resolving the B1b-N and B1b-S continuum emission}
\label{sec:alma}

The ALMA images are displayed in Fig.\ref{Fig:sources}, while Table
\ref{tab:sources} presents a summary of the observations. For each source
we show as gray contours the 32.9~GHz emission mapped with a 0.2\arcs\ beam
with the Very Large Array (VLA) as a part of the VLA Nascent Disk and
Multiplicity Survey of Perseus Protostars (VANDAM)
\citep{tobin:15a,tobin:16}. ALMA clearly resolves a compact slightly
elongated structure toward both sources that is located within a more
extended region of faint emission.

We list in Table~\ref{tab:sources} the values of flux densities at the peak
and within a 1\arcs\ radius (230~au). While both sources reach similar
intensities at large scale, the compact component in the youngest source
B1b-N is smaller and represents a smaller fraction of the total flux than
for the slightly more evolved source B1b-S. The high flux densities of the
compact components toward B1b-N and B1b-S, $> 10$\,mJy/beam, correspond to
a brightness temperature of $\sim 35$\,K, comparable to the expected gas
and dust temperatures in these sources \citep{commercon:12a}.  It is
therefore likely that the continuum emission of the compact component is
optically thick.

\begin{figure}
  \includegraphics[width=0.45\textwidth]{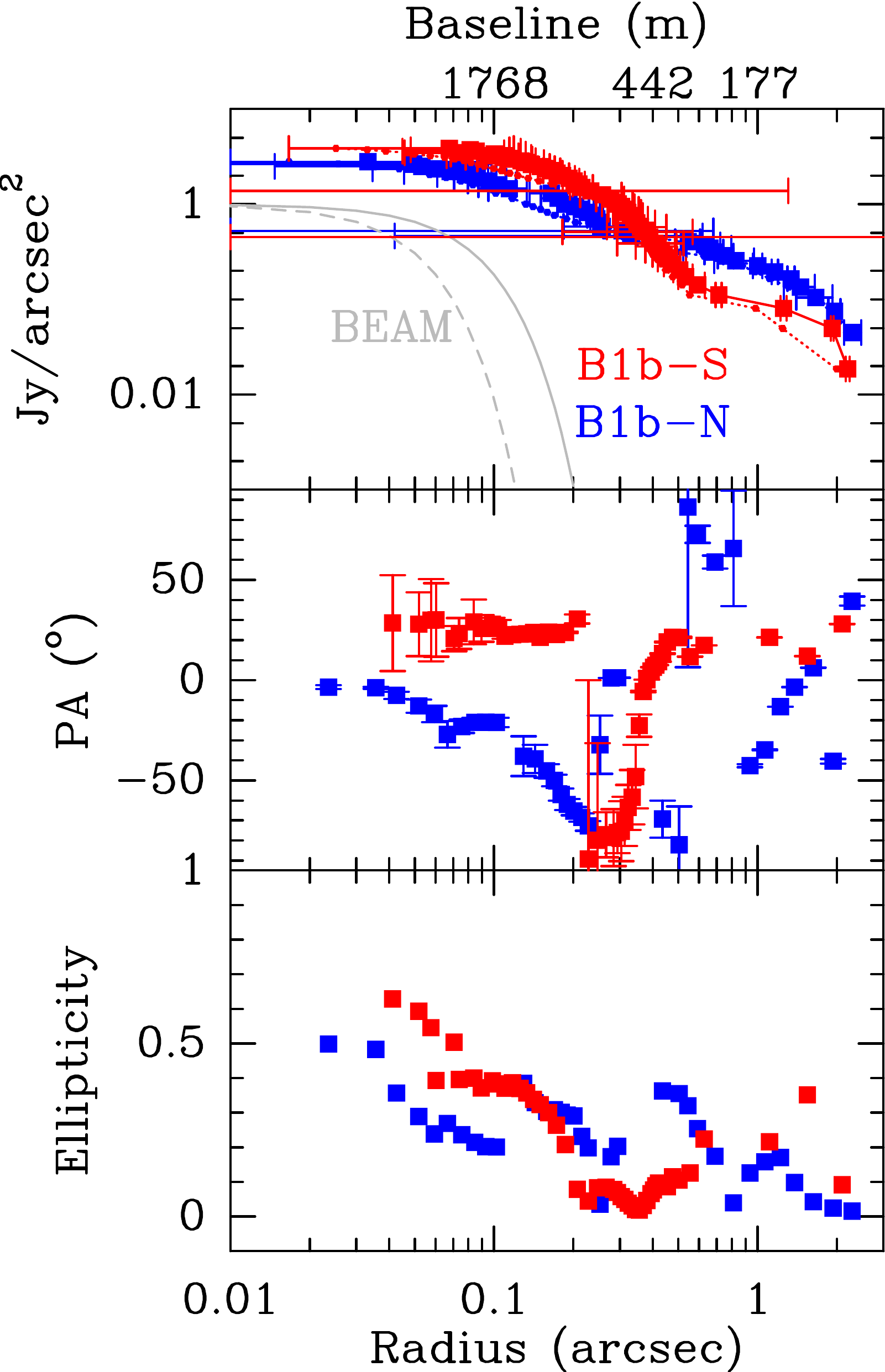}
  \caption{Fit parameters of intensity isocontours using ellipses. B1b-N
    data are shown in blue and B1b-S data in red. The top panel presents
    the variation of the major and minor axis. The middle panel shows the
    ellipse position angle, $PA$, and the bottom panel the ellipticity, $e
    = 1-R_{min}/R_{maj}$.  The beam profiles are displayed in the top
    panel. The beam position angles and ellipticities are 1.1$^\circ$ and
    0.2 for B1b-N and 16$^\circ$ and 0.21 for B1b-S.  The upper scale
    displays the correspondence between antenna baseline and angular scale.}
  \label{fig:ell}
\end{figure}

To obtain further insight into the spatial distribution of the emission, we
fit the flux density isocontours with ellipses centered on the protostar
positions, with the position angle ($PA$) and ellipticity as free
parameters.  These fits are illustrated in Figure \ref{fig:ell} with B1b-N
data in blue and B1b-S data in red. We also used simple Gaussian profiles
to determine the source sizes and shapes, and compared these estimates with
the core sizes obtained by fitting the circularly averaged intensity
profiles as described in Appendix \ref{sec:inv}.  Table \ref{tab:fit} lists
the sizes and PAs of the central sources. We obtain a FWHM of about 0.4
\arcs, corresponding to $\sim 90$~au at the Perseus distance.  The sizes of
the compact component for both sources correspond to the size of the
centimeter continuum emission, as illustrated in Fig \ref{Fig:sources}. The
similarity between the submillimeter and centimeter images, and the low
spectral index $S_{32}^{350}(0.3'')$ between 32 and 350\,GHz of 2.4 and 2.8
for B1b-N and B1b-S, respectively, confirm the high opacity of the 350~GHz
emission. It is remarkable that B1b-N is weaker than B1b-S at wavelengths
shorter than about 3~mm, but becomes stronger for longer wavelengths
\citep{gerin:15}. This behavior is explained by a larger part of the total
flux coming from the compact source at long wavelengths and the shallower
spectral index of this compact component as compared with that of the
extended envelope. Such a shallow spectral index down to at least 32~GHz is
consistent with either very high opacities even at 32~GHz or the presence
of large grains with a significantly higher emissivity at 32~GHz than
standard interstellar grains.  Studies of the submillimeter \citep{chen:16}
and mid-infrared emission \citep{lefevre:14} have shown that large dust
grains are present in Barnard~1b, as testified by the high portion of
``transitional $\beta$''\footnote{$\beta$ is the spectral index of the dust
  grain emissivity at submillimeter wavelengths.}  values, and the
coreshine phenomenon. With a standard grain population, the dust opacity at
350~GHz is a factor 50 -- 100 higher than at 32~GHz.  High opacity at
32~GHz therefore requires extreme column densities, of N(H$_2$) $\sim
10^{27}$ cm$^{-2}$, and mean H$_2$ densities higher than $\sim
10^{12}$~cm$^{-3}$ at the 0.2\arcsec scale.  We therefore propose that both
phenomena contribute to the flattening of the spectral index, the high
opacity of the continuum emission at 350~GHz and the presence of large dust
grains.

\subsection{Analysis of intensity, density, and temperature radial
  distributions}
\label{sec:analysis}

\begin{figure}
  \includegraphics[width=0.45\textwidth]{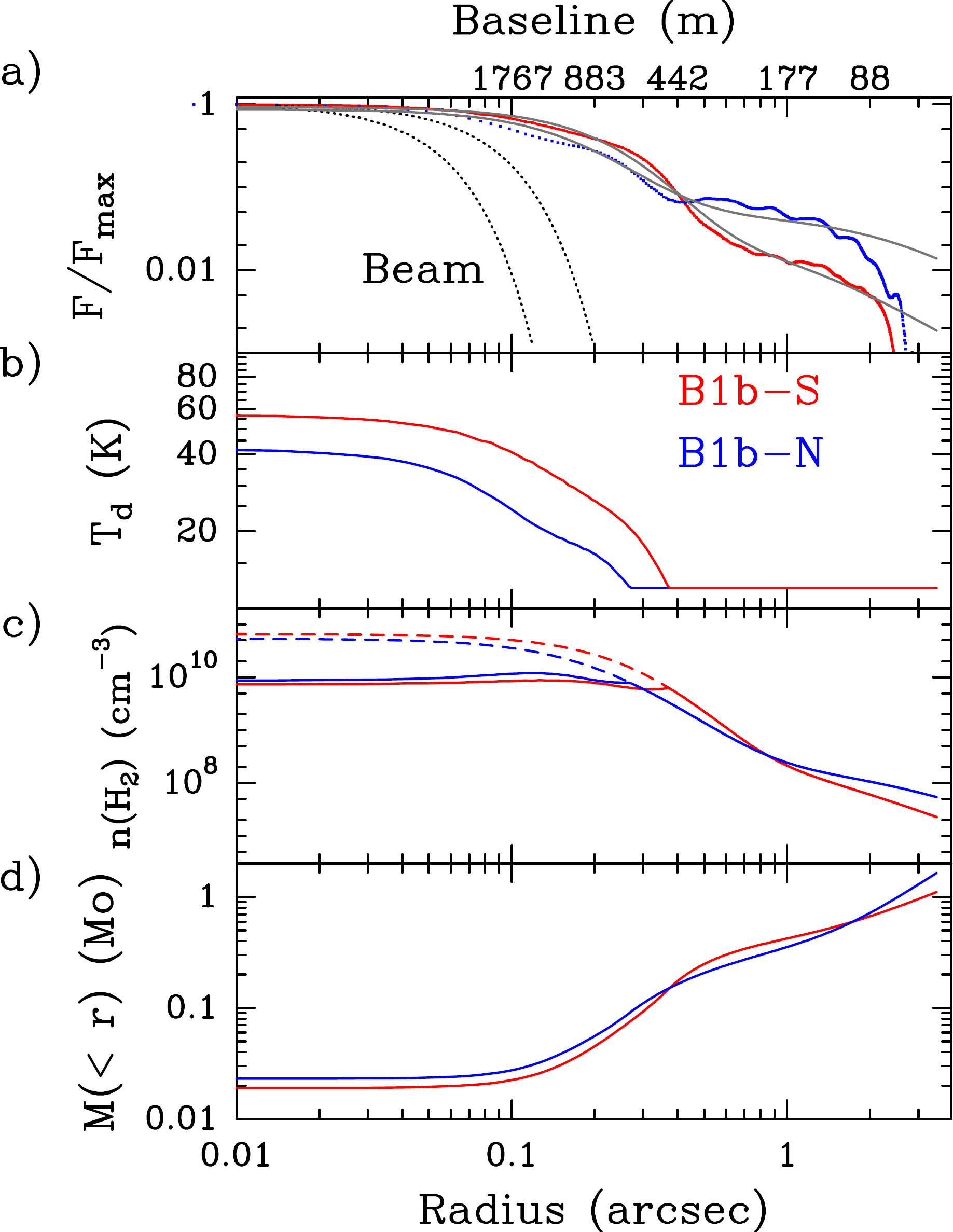}
  \caption{ a) Distribution of the circularly averaged specific intensity
    normalized to the peak value as a function of radius toward B1b-S and
    B1b-N. The gray lines show the analytical fits and the dotted lines the
    beam profiles. The upper scale presents the baseline in meters
    corresponding to the angular scale. Data for B1b-N are displayed in
    blue and data for B1b-S in red.  b) Dust temperature ($T_d$) radial
    distribution, assuming that the 350~GHz emission is optically thick,
    with a minimum value set to 12~K.  c) Molecular hydrogen density, using
    the derived $T_d$ and assuming optically thin emission. This represents
    a lower limit to the real density. The dashed lines show the densities,
    assuming a constant value of the dust temperature.  d) Cumulative mass
    distribution obtained by combining the information at 350 and 32~GHz. }
  \label{fig:rad}
\end{figure}

Figure \ref{fig:rad} presents the observed distributions of the circularly
averaged specific intensity as a function of the angular radius for B1b-N
and B1b-S and their associated fits in the top panel, together with the
variations of dust temperature, molecular hydrogen density, and the
cumulative gas masses in the other panels.  While the dust continuum
emission from the compact component (disk) is probably optically thick, the
sharp decrease in flux densities outside the central component indicates
that the opacity of the dust thermal emission drops. The assumption of
optically thin emission is therefore likely valid for the envelope
region. Using the procedure outlined in Appendix \ref{sec:inv}, we derived
the radial distribution of the gas density. This assumes spherical
symmetry, that the gas is molecular, and that the dust has constant optical
properties, $\kappa = 0.01(\frac{\nu}{230~\mathrm{GHz}})^{1.8}$\,
cm$^2$g$^{-1}$, a value commonly used for cold and dense cores, from
\cite{oh94}, which is close to the composite aggregate disk opacities
computed by \citet{semenov:03}.  A further assumption is the good thermal
coupling of the dust and gas, which leads to equal gas and dust
temperatures, supported by the high densities encountered in the cores
\citep[$n(\mathrm{H}_2) > 10^5$ cm$^{-3}$,][]{daniel:13}.  Finally, we used
a simple equation of state to relate the dust temperature and the gas
density: {\sl i)} the gas is isothermal and the kinetic temperature $T_0$
is 12~K \citep[derived by][from NH$_3$]{lis:10} for all densities lower
than a density threshold $n_0(\rm H_2) = 10^{10}$\,cm$^{-3}$; {\sl ii)} the
gas follows a polytropic equation of state for densities higher than
$n_0(\rm H_2)$, with an exponent $\gamma$, implying $T(r) = T_0
(\frac{n(r)}{n_0(\rm H_2)})^{\gamma-1}$.  We have chosen $\gamma = 1.6$,
adapted for moderately warm molecular gas.  For the ``disk region'' within
a radius of 0.3\arcs, we derived the dust temperature directly from the
brightness temperatures.  As the 350~GHz emission is optically thick, we
used the 32~GHz flux densities to determine the gas masses, assuming an
emissivity of $\kappa = 0.006$\, cm$^2$g$^{-1}$ at 1~cm, adapted to a
population of large grains with sizes reaching 3\,mm \citep{tazzari:16},
and using the temperature determined from the 350~GHz data.  The total
masses were derived by combining the determinations at 32~GHz and at
350~GHz.

The parameters used for the fit and the resulting physical conditions are
given in Table \ref{tab:fit}.  For both sources, we used a two-component
fit with a central component with a steep flux density profile ($r^{-4}$ or
$r^{-6}$) and a more extended envelope with flux decreasing as
$r^{-2}$. Because of the lack of short spacings, some flux may be missing
at scales larger than $\sim 2$~\arcsec. Therefore the parameters of the
second component are not well constrained.  The exponent and size scale
should not be considered as having a physical meaning, but only as a
convenient mathematical description of the observed data.  Assuming
optically thin 350~GHz emission, the maximum H$_2$ density reached in the
central region is about 10$^{10}$ cm$^{-3}$ for both sources. The true
densities are expected to be significantly higher.  The peak dust
temperature is 40 and 60\,K, somewhat higher than the mean temperature used
for the SED fit by \citet{hirano:14} (16 and 18\,K for B1b-N and B1b-S),
but still consistent with faint emission at wavelengths shorter than
70$\mu$m.  For both sources, the warm region where the dust temperature is
higher than 12\,K is fairly small, less than 0.3\arcs in radius.  The
cumulative mass distributions show that the sources may still be accreting,
since a large part of the core mass is located at radii larger than
0.3\arcs. B1b-S is more condensed and presents a steeper increase in mass
with radius than B1b-N, supporting its more evolved classification
(established from the larger extent of its molecular outflow and warmer
SED; \citet{gerin:15}).  The mass of the compact component is at least
$0.1$~M$\sun$ for both B1b-N and B1b-S, given our hypothesis on the dust
grain emissivity at 32~GHz.

\begin{table*}
  \caption{Analytical fits and physical structure}
  \label{tab:fit}
  \begin{tabular}{lcccc}
    \hline
    \hline
    & \multicolumn{2}{c}{B1b-S} & \multicolumn{2}{c}{B1b-N}   \\
    \hline
    $Size$$^a$ (\arcs) &  \multicolumn{2}{c}{$0.45 \times 0.43 \pm 0.05$} &  \multicolumn{2}{c}{$0.42 \pm 0.4$} \\
    $PA$$^a$ ($^\circ$) & \multicolumn{2}{c}{$25\pm 10$} &   \multicolumn{2}{c}{$170\pm 10$} \\
    \hline
    $F_0$, $F_1$$^b$ (mJy/beam) & $66 \pm 2$ & $1.1\pm 0.4$ & $15.3 \pm 0.6$ & $0.8 \pm 0.1$    \\
    $r_0$, $r_1$$^b$ (\arcs) & $0.34 \pm 0.005$ & $1.06 \pm 0.22$ &  $0.21 \pm 0.005$ & $2.3 \pm 0.5$\\
    $\alpha_0$, $\alpha_1$$^b$ & $6$ & $2$ &  $4$  & $2$\\
    \hline
    $n_0$$^c$ (H$_2$ cm$^{-3}$) & \multicolumn{2}{c}{$0.9 \times 10^{10}$} &   \multicolumn{2}{c}{$1.0 \times 10^{10}$}\\
    $\rho_g$$^c$ (g\,cm$^{-3}$) & \multicolumn{2}{c}{$3.9 \times 10^{-14}$} &   \multicolumn{2}{c}{$4.3 \times 10^{-14}$}\\
    $T_c$$^d$ (K) & \multicolumn{2}{c}{56} &  \multicolumn{2}{c}{41} \\
    $T_{0}$ (K) & \multicolumn{2}{c}{12} &  \multicolumn{2}{c}{12}\\
    $M(r<0.3'')$ (M$_\sun$) & \multicolumn{2}{c}{0.093}  &  \multicolumn{2}{c}{0.11}\\
    $M(r<1.0'')$ (M$_\sun$) & \multicolumn{2}{c}{0.42}  &  \multicolumn{2}{c}{0.36}\\
    $M(r<2.0'')$ (M$_\sun$) & \multicolumn{2}{c}{0.67}  &  \multicolumn{2}{c}{0.72}\\
    \hline
  \end{tabular}
  \tablefoot{$^a$ From Gaussian fitting of the ALMA images.
    $^b$The fitting formula of the circularly averaged specific distribution for B1b-S  and  for B1b-N uses two modified power laws  :
    $F(r)= \frac{F_0}{(1+\frac{r^2}{r_0^2})^\frac{\alpha_0}{2}} + \frac{F_1}{(1+\frac{r^2}{r_1^2})\frac{\alpha_1}{2}}$.
    For isothermal optically thin dust emission, the slope of the
    density distribution is $\alpha + 1$.
    $^c$ Assuming optically thin emission at 350~GHz.
    $^d$ Deduced from the specific intensity assuming optically thick emission. }
\end{table*}

\begin{figure}
  \includegraphics[width=0.45\textwidth]{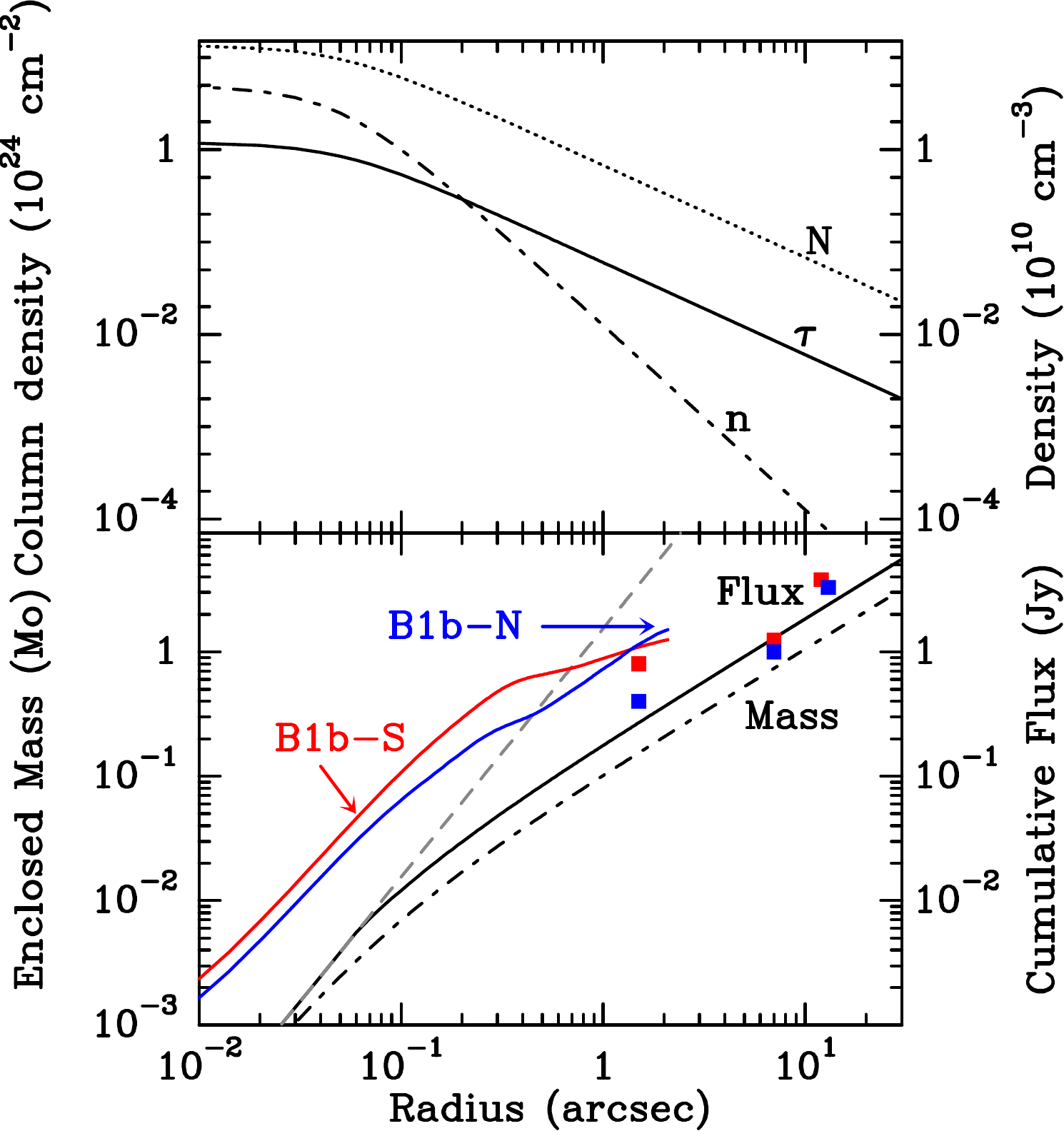}
  \caption{ Top panel: variation with radius of the molecular hydrogen
    density, $n$ (dash-dotted line), the column density, $N$ (dotted line),
    and the resulting continuum opacity at 350 GHz, $\tau$ (solid line) for
    a simple spherical isothermal (12~K) core model. Bottom panel:
    variation of the gas mass (dash-dotted line) and 350 GHz flux (solid
    line) as a function of radius for the model displayed in the top
    panel. The dashed gray line shows the expected cumulative flux
    distribution for a pure blackbody (optically thick emission) at the
    same temperature. The observed data toward B1b-N and B1b-S are
    displayed in blue and red, respectively, including the ALMA data from
    0.05" to 3" and JCMT and SMA data between 3" and 12"
    \citep{pezzuto:12,hirano:14,li:17}. }
  \label{fig:sphere}
\end{figure}

To check whether the compact component detected with ALMA could simply
represent the inner region of a protostellar envelope, we compared the
cumulative flux distribution of a simple isothermal spherical envelope
model with that of B1b-N and B1b-S. We used a density profile $n({\rm H_2})
= \frac{5 \times 10^{10} {\rm cm^{-3}} }{1+(r/0.05")^2}$ , where $r$ is the
radius in arcseconds, a uniform temperature of 12~K, and the dust
emissivity law described above. These parameters lead to a predicted flux
in a 20" beam of about 2~Jy, comparable to the measured flux density of
B1b-N and B1b-S with SCUBA, about 3~Jy \citep{pezzuto:12,hirano:14,li:17},
and a high opacity of the central region within a radius of 0.1 arcsec, as
illustrated in Fig. \ref{fig:sphere}.

However, this model presents a cumulative flux distribution very different
from that of B1b-N and B1b-S, which have a more spatially extended region
of high opacity and a higher temperature, supporting the association of the
compact component detected with ALMA with a disk. To further understand the
nature of this compact component, we have used numerical simulations of
collapsing cores.

\section{3D collapse model and comparison with the ALMA data}
\subsection{Description of the model}
\label{sec:3Dmodel}

\begin{figure}[t]
  \includegraphics[width=0.45\textwidth]{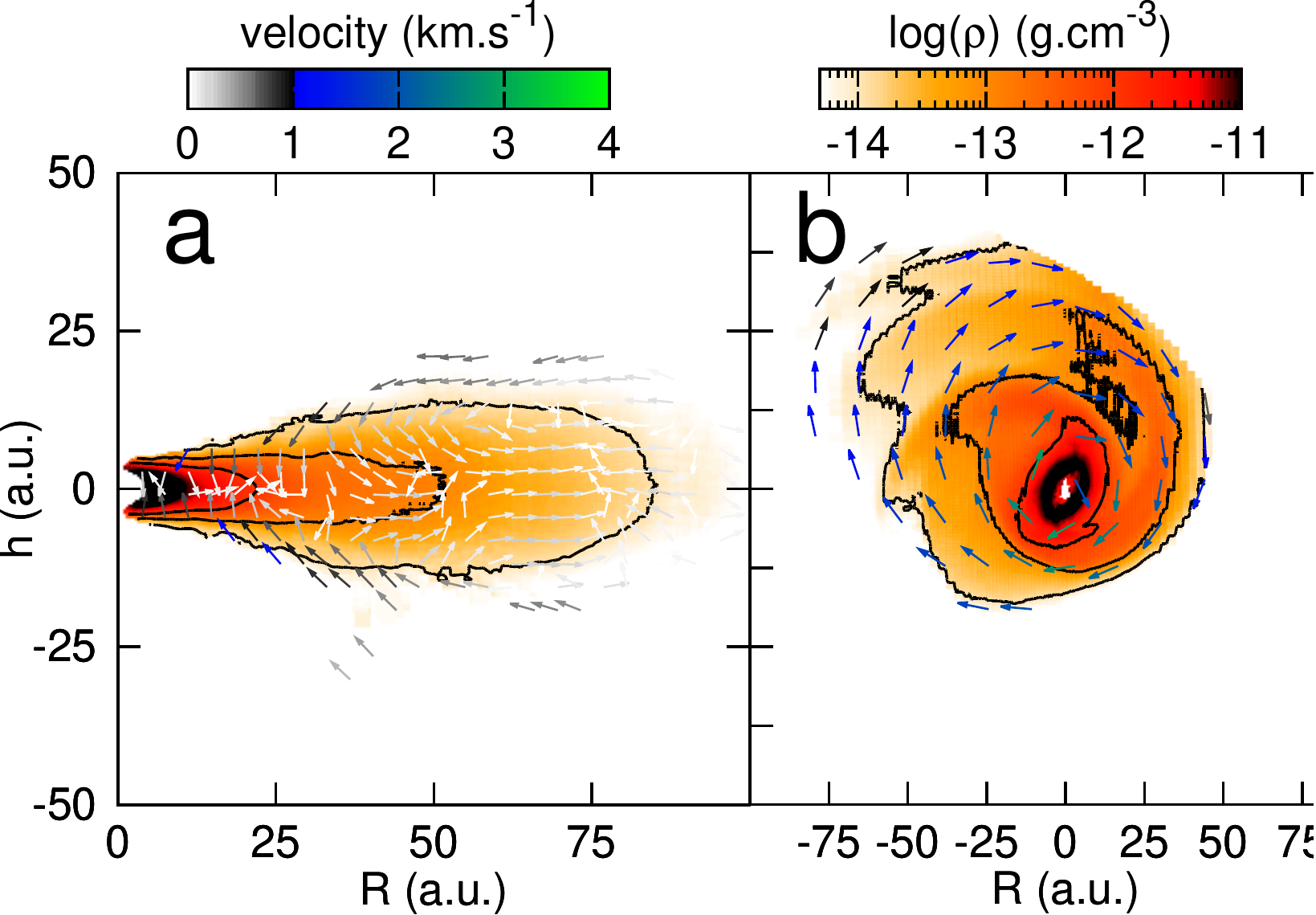}
  \caption{Cross section of the 3D numerical simulations showing the
    structure of the protostellar disk. The colors show the density field
    and the arrows the velocity vectors. Panel (a) shows an edge-on view
    (azimuthal mean) and panel (b) a face-on view of all cells that satisfy
    the criteria for a rotationally supported disk \citep{joos:12}.  }
  \label{Fig:disk}
\end{figure}

In this section, we compare our ALMA data with simulated observations
obtained by post-processing a 3D numerical simulation from
\cite{hennebelle:16}. We have used the \ttfamily{RAMSES}\rm~code
\citep{teyssier:02}, which integrates the MHD equations using a constrained
transport scheme \citep{fromang:06,teyssier:06} and accounts for the
ambipolar diffusion \citep{masson:12}. The model is almost identical to the
model used in \cite{masson:15}, except that we account for initial
turbulent velocity fluctuations instead of a coherent global solid body
rotation \citep[e.g., ][]{joos:13}.  The model consists of a collapsing
1M$_\odot$ magnetized and turbulent dense core of radius $R_0=2500$~au. The
initial dense core density and temperature are uniform, $\rho_0=9.4\times
10^{-18}$ g~cm$^{-3}$ corresponding to $n(\rm H_2) = 2.4 \times
10^6$\,cm$^{-3}$ and $T=10$~K. The corresponding ratio between the thermal
energy and the gravitational energy is $\alpha_\mathrm{th}=0.25$. The
initial mass-to-flux ratio parameter is $\mu=2$ (strongly magnetized
model). The initial turbulent velocity fluctuations are introduced
following \cite{joos:12} (see their Sect. 2.2). The velocity fluctuations
are scaled to match an initial Mach number $\mathcal{M}=1.2$. The thermal
behavior of the collapsing material is reproduced using a barotropic
equation of state. The initial resolution is $32^3$ and the mesh is refined
to always describe the Jeans length with at least eight points, down to a
resolution of $\sim 0.6$~au.

\begin{figure}
  \includegraphics[width=8cm]{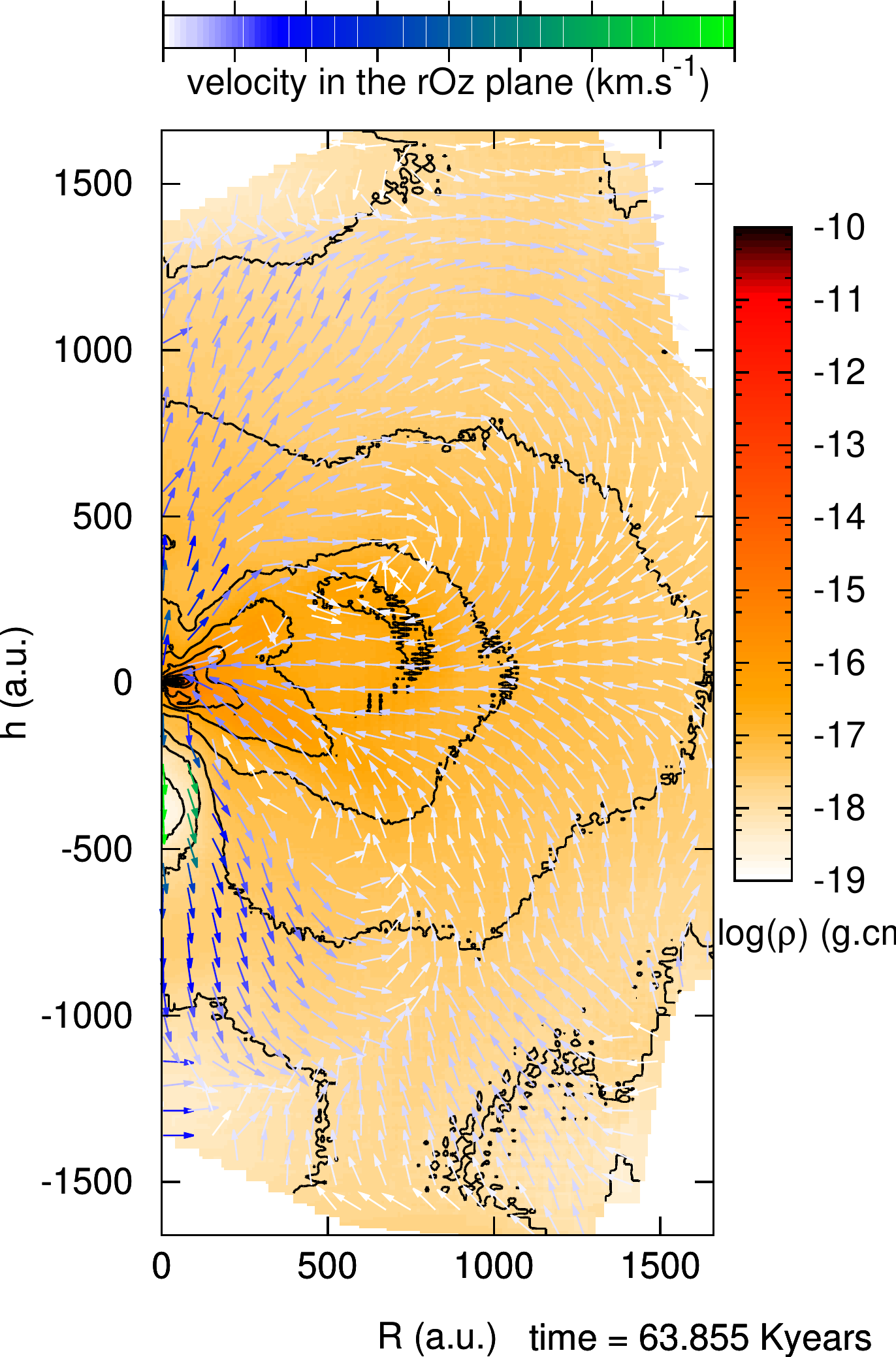}
  \caption{Azimuthal average of the density and velocity fields of the
    collapsing core at the same time as in Fig. \ref{Fig:disk}. The outflow
    extends up to $\sim 1100$ au. The velocities range from 0 to 4 \kms,
    the density from 10$^{-19}$ to 10$^{-10}$~g\,cm$^{-3}$, or from $2.4
    \times 10^5$ to $2.4 \times 10^{13}$ H$_2$\,cm$^{-3}$.}
  \label{Fig:outflow}
\end{figure}

Figure \ref{Fig:disk} shows the protostellar disk structure in the
models$~6.3$~kyr after the FHSC formation. The rotationally supported disk
is identified following the criteria derived in \cite{joos:12}. The disk
extends up to $\sim 75$~au, and $98\%$ of its mass is contained within a
radius $< 60$ au. The FHSC mass is $0.09$ M$_\odot$ and the disk mass
$0.07$ M$_\odot$. Figure \ref{Fig:outflow} shows an azimuthal average of
the density and velocity fields around the rotation axis. The outflow
extends up to $\sim1100$ au.
 
\subsection{Synthetic observations}
\label{Sec:synthobs}

We post-processed the simulation using the the
\ttfamily{RADMC-3D}\rm~radiative transfer code \citep{dullemond:12} to
compute the expected 350~GHz emission. Following
\cite{commercon:12a,commercon:12b}, we assumed that the gas and dust
temperature are perfectly coupled and that the dust-to-gas ratio is uniform
and equal to 1\%. We used the dust opacity from \cite{semenov:03} for the
homogeneous spheres model (with Fe/Fe+Mg=0.3 "normal" silicate
composition).  We used the Barnard~1 distance and performed the computation
for a range of inclinations $i$ between face-on to edge-on. The emission
structure is not very sensitive to the inclination between about 20$^\circ$
and 60$^\circ$.

\subsection{Comparison of models and observations}
\label{sec:model}

\begin{figure}
  \includegraphics[width=0.45\textwidth]{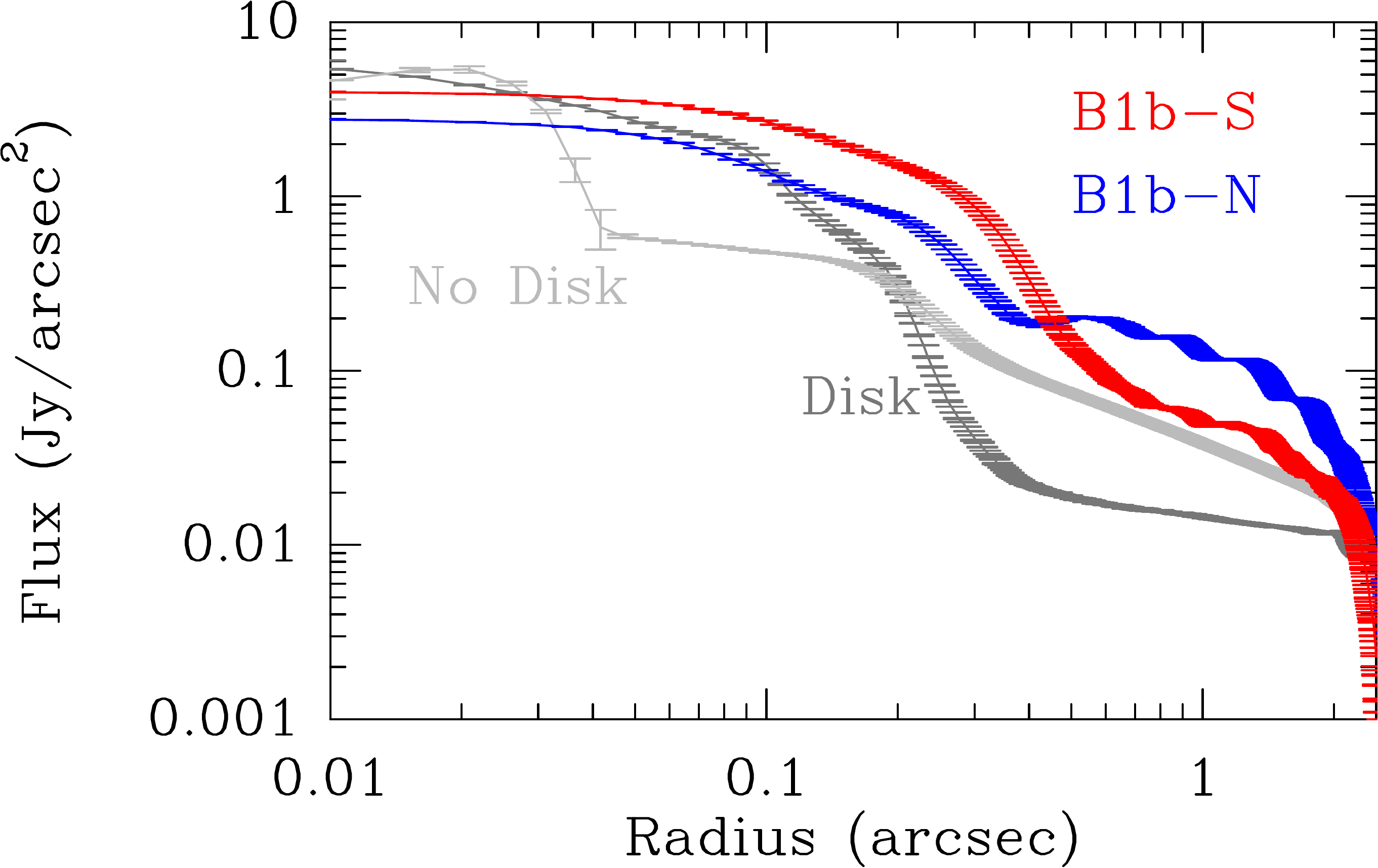}
  \caption{Comparison of the radially averaged flux distribution for B1b-S
    (red), B1b-N (blue), and two different epochs of the MHD collapse
    model. The light gray curve labeled ``No Disk'' shows an early epoch
    before the formation of the FHSC, while the darker gray curve labeled
    ``Disk'' shows the later time illustrated in Figs. \ref{Fig:disk} and
    \ref{Fig:outflow}. }
  \label{fig:comparison}
\end{figure}

Figure \ref{fig:comparison} presents the comparison of the radially
averaged flux toward B1b-N and B1b-S with the predictions obtained from the
model at two epochs.  We used the views with $i = 20^\circ$ for two
different epochs: an early stage of the collapse, $400$ yr before the
formation of the FHSC, when the gas structure remains close to that of an
isothermal sphere and no disk is formed yet, and a late stage ($6.3 \times
10^3$ yr after the formation of the FHSC) when a rotationally supported
disk has already formed, as illustrated in Fig. \ref{Fig:disk} and
\ref{Fig:outflow}.  The emission structure of the late-stage model is
similar to that of the B1b protostars: both the disk radius and the
predicted flux density radial distribution agree within better than a
factor of two with the observations. The early-stage model presents a very
different structure, with a more localized emission peak of about 0.04
\arcs, and a rapidly decreasing flux significantly lower than that detected
toward B1b-N and B1b-S, as expected from the similarity between this early
stage and the simple model shown in Fig. \ref{fig:sphere}.  The difference
of flux radial distributions between the two epochs stems from the
different temperature and density distributions. In particular, the high
surface brightness region is more localized when no disk is formed (0.04"
vs 0.2") because the very high density region (n(H$_2$) $> 10^9$~cm$^{-3}$)
is reduced.  The excellent agreement of the observed 350~GHz emission
profile with the model profile confirms the high opacity of the dust
emission and indicates that the dust temperature remains moderate.

To further explore the similarity of the model with the B1b-N data, we now
consider the outflow spatial extend and maximum velocity, which are
illustrated in Fig. \ref{Fig:outflow}. The values predicted by the model
are fairly similar to those of B1b-N \citep{gerin:15}.  The good
correspondence of the central sources' size and shape, combined with their
detection at 32~GHz, the extremely high molecular hydrogen densities and
their estimated masses, support the identification of these sources with
rotationally supported disks. This identification remains tentative in the
absence of kinematic data, however.  The sizes of the tentative disks are
within a factor of a few from the predictions using the analytical
expression derived by \citet{hennebelle:16}.  Although fairly uncertain
because we lack information about the grain emissivity law, especially when
large grains are present \citep{dunham:14}, the disk masses and
temperatures are consistent with theoretical predictions of low-mass star
collapse models
\citep{masunaga:00,machida:10,dunham:14,masson:15}. \citet{zhao:16} have
shown that the removal of small grains can favor the formation of disks by
reducing the coupling of the matter with the magnetic field, hence the
efficiency of magnetic braking. The presence of large dust grains in
Barnard~1b argues for a lower fraction of small dust grains in this core,
which may have favored the formation of disks.  We conclude that although
no kinematic information is available so far, the sizes ($\sim 50$~au) and
brightnesses are consistent with those of rotationally supported disks
formed during the collapse of dense cores in current MHD simulations. B1b-N
and B1b-S thus appear to be excellent objects for testing our understanding
of disk formation during low-mass protostar formation, which deserve
further studies of their kinematics and spatial structure.
  
\begin{acknowledgements}
  This paper makes use of the following ALMA data:
  ADS/JAO.ALMA\#2015.1.00025.S and ADS/JAO.ALMA\#2015.1.00025.S. ALMA is a
  partnership of ESO (representing its member states), NSF (USA) and NINS
  (Japan), together with NRC (Canada), NSC and ASIAA (Taiwan), and KASI
  (Republic of Korea), in cooperation with the Republic of Chile. The Joint
  ALMA Observatory is operated by ESO, AUI/NRAO and NAOJ.  We acknowledge
  funding support from PCMI-CNRS/INSU, PNPS-CNRS/INSU, AS ALMA,
  Observatoire de Paris, ANR project ANR-13-BS05-0008 (IMOLABS) and the
  European Research Council (ERC Grant 610256: NaNOCOSMOS). We thank IRAM
  ARC for help with the data processing, and J. Tobin for access to the
  VANDAM images. We thank the referee for a careful reading of the
  manuscript which led to a significant improvement of our analysis.  This
  work was performed using HPC resources from GENCI-CINES (Grant
  2016-047247).
\end{acknowledgements}

\bibliographystyle{aa} %
\bibliography{aa30187}

\begin{appendix}

  \section{Calibration}
  \label{sec:cal}
  
\begin{figure}
    \includegraphics[width=4cm]{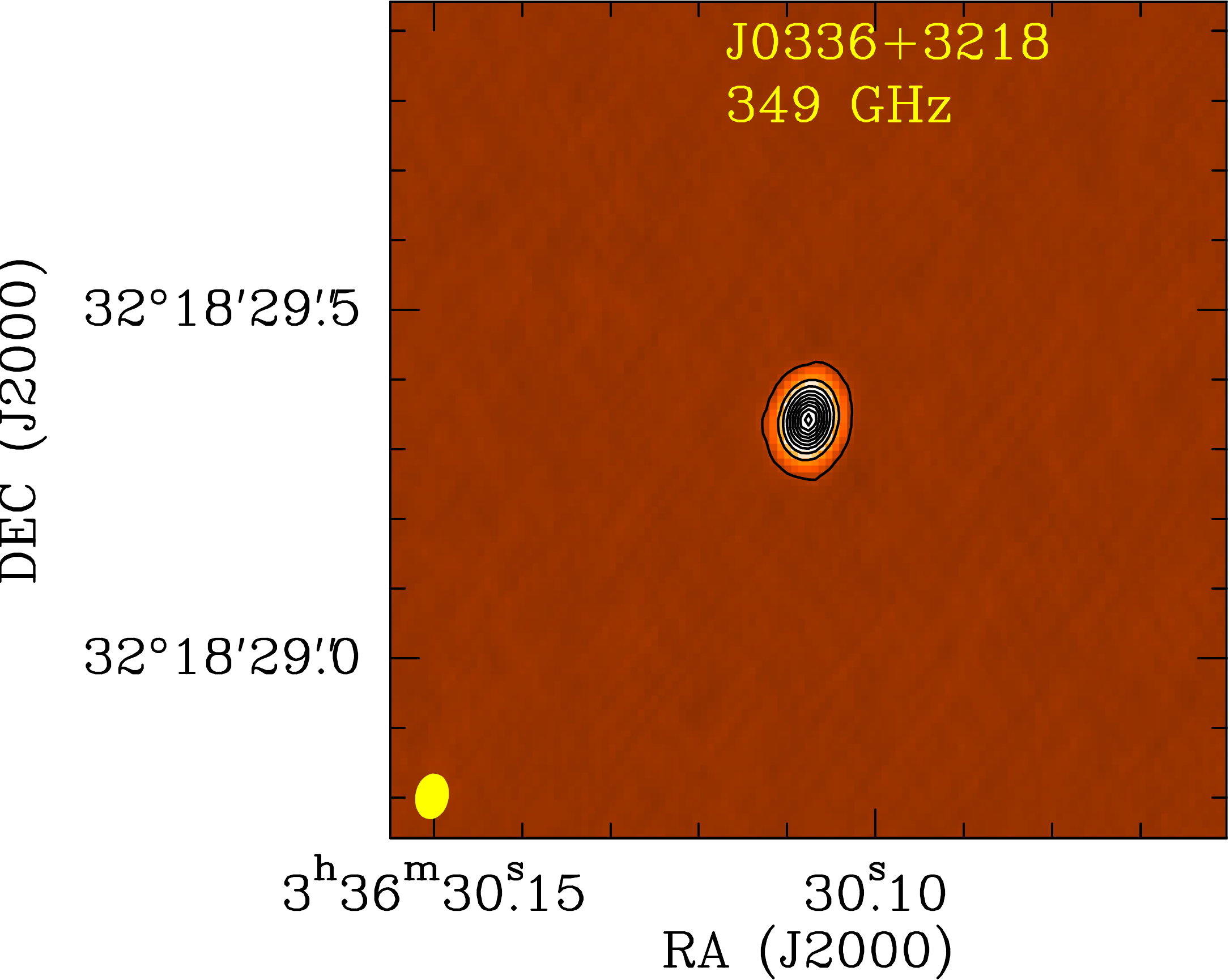}
    \hspace*{0.2cm}
    \includegraphics[width=4cm]{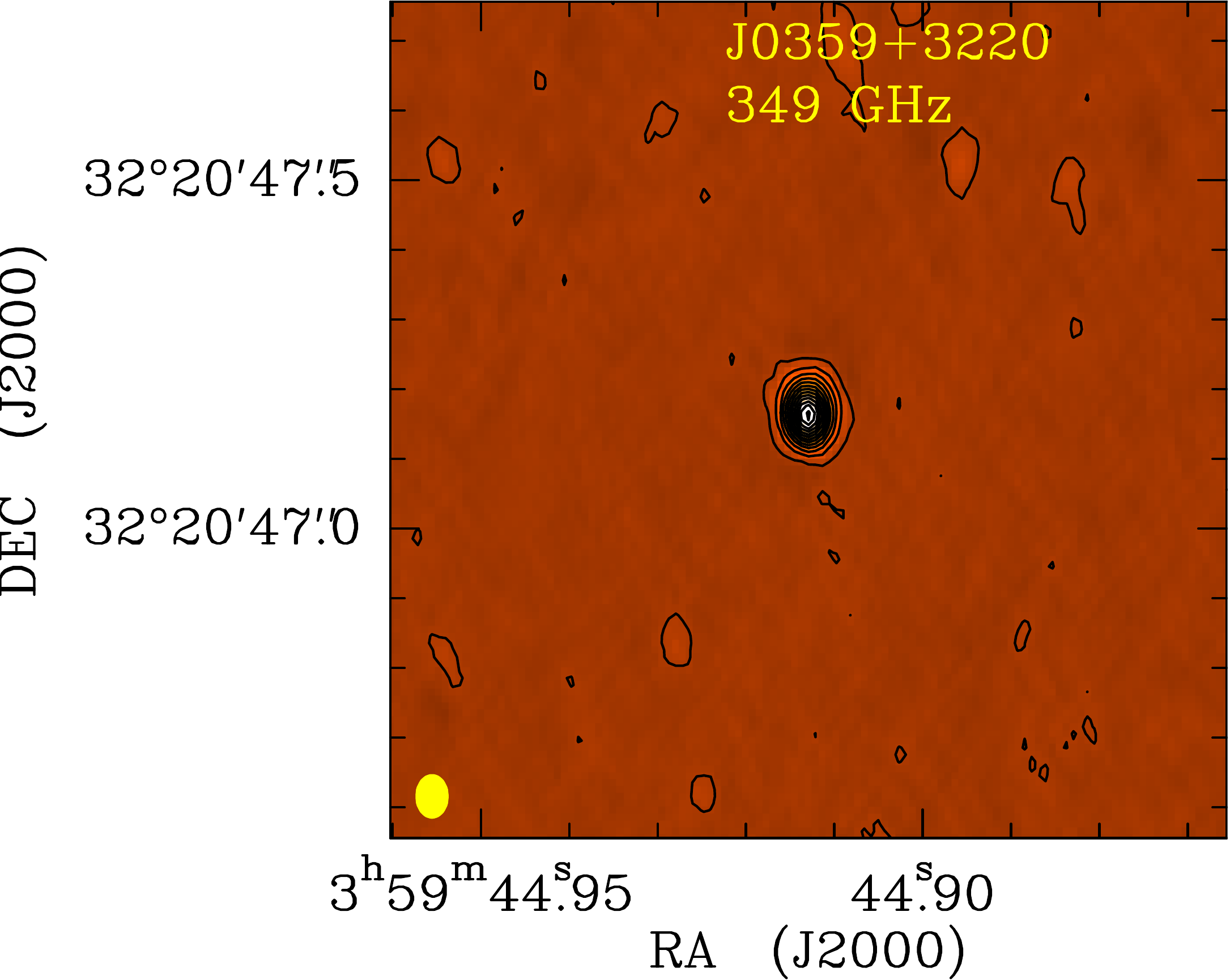}
    \caption{Images of the point-like continuum sources J0336+3218 (Left)
      and J0359+3220 (right) observed during the first observing session
      (C36-7 configuration).  The yellow ellipse shows the ALMA beam.  }
    \label{Fig:check}
  \end{figure}

  In order to check the quality of the phase-calibration transfer to the
  source targets for the long-baseline dataset, we first imaged the control
  source (J0359+3220) using the phase solution obtained on the phase
  calibrator (J0336+3218). However, the solution was only moderately
  satisfying because J0359+3220 is relatively far from the phase calibrator
  (5.7$^\circ$), and the quasar appears slightly extended. In such
  conditions, the imaging results do not reflect the real performances on
  the targets.  Therefore we performed a second test on the phase
  calibrator J0336+3218.  In these tests, we used half of the scans
  observed on the phase calibrator to obtain the calibration tables, which
  then were used to transfer the phase calibration onto the other half of
  scans. For the second half, we therefore performed phase transfer instead
  of self-calibration.  The images are acceptable for the phase-transferred
  scans, which shows that we can rely on the phase calibration of the
  targets.  We also performed the cleaning in GILDAS in the same way as for
  the B1b sources. The resulting images are displayed in
  Fig~\ref{Fig:check}.

\begin{figure}
    \centering
    \includegraphics[width=0.45\textwidth]{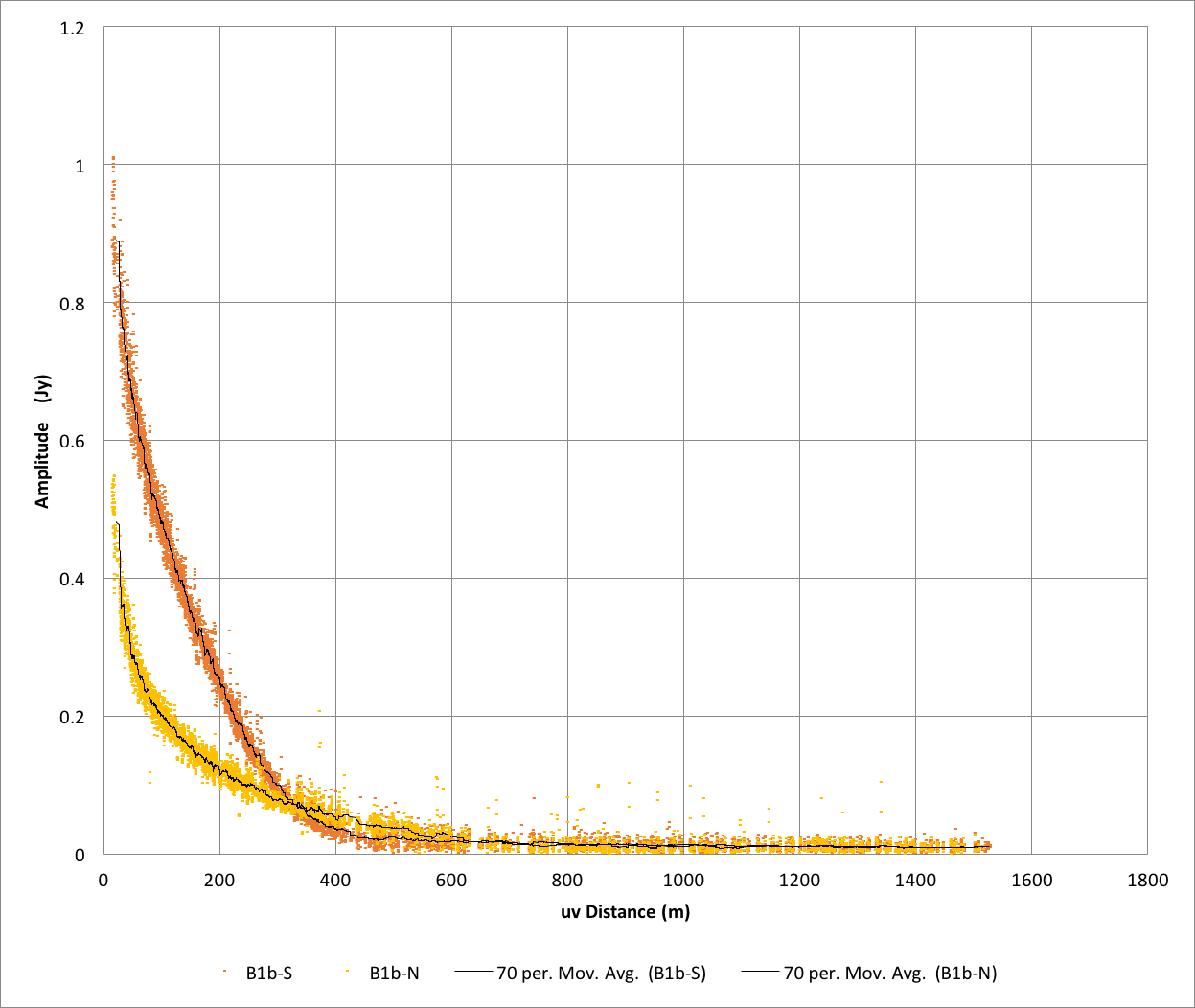}
    \caption{Amplitude of the visibility flux in Jansky versus UV distance
      in meters for the two data sets. The dots show the full sets of
      measurements, while the lines trace the moving average values using 70
      points.  B1b-S is shown in orange, B1b-N in yellow.}
    \label{fig:uv}
  \end{figure}

  Figure \ref{fig:uv} presents the amplitude versus UV distance for the
  combined data sets. The good overlap of the curves of flux versus UV
  distance in the domain from 20~m up to 600~m shows that the flux
  calibration is very good.  B1b-N and B1b-S are heavily resolved, with
  most of their fluxes on short baselines.

  \section{Inversion of radial profile}
  \label{sec:inv}

  As shown by \citet{krco:16}, it is possible to use the radial profile of
  the projected 2D image of a spherically symmetrical source to retrieve
  the information on the initial 3D radial distribution. This appendix
  describes the method we applied to the submillimeter images.  Assuming a
  spherical source of radius $R_{max}$, at a distance $D$, the flux density
  can be expressed as the integrated emission along the line of sight:
  \begin{equation}
    dF_\nu (r) = 2 \Omega \int_r^{R_{max}} \kappa_\nu n(z)B_\nu(T(z)) \frac{z}{\sqrt{z^2-r^2}}dz, 
  \end{equation}
  where $n$ is the dust density, $\kappa_\nu$ is the dust emissivity, and
  $\Omega$ the beam size.  The total flux density is then
  \begin{equation}
    F_\nu^{Tot} = \frac{1}{D^2}\int_0^{R_{max}} F_\nu(r)2\pi r dr.
  \end{equation}
  Since the density and temperature only depend on the radius, it is
  possible to inverse the radial distribution of the flux density to derive
  the product of the density and Planck function $n(r)B_\nu(T(r))$.
  Inversion methods are rather sensitive to the noise level. We therefore
  chose another approach and fit the radial distributions of the flux
  density with an analytical radial profiles.  We used the analytical
  solution of the inversion to determine $n(r)B_\nu(T(r))$.  The two
  variables can be separated in a following step, using a simple assumption
  for the temperature profile as described in Section \ref{sec:analysis}.
  We used a simple attenuated power-law profile for the radial distribution
  of the flux density \citep[see, e.g.,][]{krco:16},
  \begin{equation}
    F(r) = \frac{F_0}{(1+\frac{r^2}{r_0^2})^{\frac{\alpha}{2} } },
  \end{equation}
  which can be easily inverted to yield
  \begin{equation}
    n(r)B_{\nu}(T(r)) = \frac{F_0}{\kappa_\nu \Omega r_0D}
    \frac{\Gamma(\frac{\alpha+1}{2})}{\Gamma(\frac{\alpha}{2})\sqrt{\pi}}(1+\frac{r^2}{r_0^2})^{-\frac{1+\alpha}{2}},
  \end{equation}
  where $\Gamma(x)=(x-1)!$ with $x$ an integer number $\geq 2$.
 
  Developing $B_\nu(T(r))$ yields in the Rayleigh-Jeans limit
  \begin{equation}
    B_\nu(T(r)) = \frac{2h\nu^3}{c^2}\frac{1}{e^\frac{h\nu}{k_BT(r)}-1} = \frac{2k_b\nu^2}{c^2}T(r).
  \end{equation}
  We note that at 350 GHz, the Rayleigh-Jeans limit is only valid for
  $T\geq 35$~K, hence it is better to use the actual Planck function.  The
  fitting parameters for both sources are displayed in Table \ref{tab:fit}.

\end{appendix}

\end{document}